\documentclass[fleqn,usenatbib]{mnras}
\usepackage[T1]{fontenc}
\usepackage{ae,aecompl}
\usepackage{graphicx}	
\usepackage{amsmath}	
\usepackage{amssymb}	
\title[Peculiar motion from Hubble diagram of SNe Ia]{Peculiar motion of Solar system from the Hubble diagram of supernovae Ia and its implications for cosmology}
\author[A. K. Singal]{Ashok K. Singal\thanks{E-mail: ashokkumar.singal@gmail.com}\\
{Astronomy and Astrophysics Division, Physical Research Laboratory, 
Navrangpura, Ahmedabad - 380009, India}}
\date{Accepted XXX. Received YYY; in original form ZZZ}
\pubyear{2019}
\begin{document}
\label{firstpage}
\pagerange{\pageref{firstpage}--\pageref{lastpage}}
\maketitle
\begin{abstract}
Peculiar motion of the solar system, determined from the dipole anisotropy in the Cosmic Microwave Background Radiation (CMBR), has given a velocity $370$ km s$^{-1}$ along RA$=168^{\circ}$, Dec$=-7^{\circ}$. 
Subsequent peculiar motion determinations from the number counts, sky brightness or redshift dipoles observed in large samples of distant radio galaxies and quasars yielded peculiar velocities two to ten times larger than CMBR, though in all cases the directions matched with the CMBR dipole.  
Here we introduce a novel technique for determining the peculiar motion from the magnitude-redshift ($m_{\rm B}-z$) Hubble diagram of Type Ia Supernovae (SN Ia), one of the best standard candles available. We find a peculiar velocity $1.6\pm 0.5 \times 10^3$ km s$^{-1}$,  larger than the CMBR value roughly by a factor of four, along RA$=173^{\circ}\pm12^{\circ}$, Dec$=10^{\circ}\pm9^{\circ}$, the direction being within $\stackrel{<}{_{\sim}}2\sigma$ of the CMBR dipole.
Since a genuine solar motion would not depend upon the method or the dataset  employed, large discrepancies seen among various dipole amplitudes could imply that these dipoles, including the CMBR one, might not pertain to observer's  peculiar motion.
However, a common direction for various dipoles might indicate a preferred direction in the universe, implying an intrinsic anisotropy, in violation of the cosmological principle, a cornerstone of the modern cosmology.
\end{abstract} 
\maketitle
\begin{keywords}
methods: miscellaneous -- supernovae: general  --  cosmic background radiation -- cosmological parameters -- large-scale structure of Universe -- cosmology: miscellaneous
\end{keywords}
\section{INTRODUCTION}
Exploiting SNe~Ia as standard candles, cosmological parameters have been derived, which suggested a Universe with accelerating expansion rate (Riess et al. 1999; Perlmutter et al. 1999; Betoule et al. 2014; Jones et al. 2018; Scolnic et al. 2018). 
Some reservations have recently been expressed on these results
(Nilson, Guffanti \& Sarkar 2016; Colin et al. 2019), 
but these have been refuted elsewhere (Rubin \& Hayden 2016; Rubin \& Heitlauf 2020).
One important element in these studies is the estimate of the local bulk flow (Mohayaee, Rameez \& Sarkar 2020). For instance, no clinching evidence for such bulk flow from type Ia supernovae has been found (Huterer, Shafer \& Schmidt 2015). On the other hand, bulk flows  $v_{\rm bf}\sim$ 170 to 540 km s$^{-1}$ in the nearby universe ($z\stackrel{<}{_{\sim}}0.06$) have also been reported (Weyant et al. 2011; Turnbull, et al. 2012; Mathews, et al. 2016; Boruah, Hudson \& Lavaux 2020).
In all these studies, the observed heliocentric redshifts and magnitudes of SNe were corrected for the peculiar velocity of the observer, derived from the CMBR dipole 
to be 370 km s$^{-1}$ in the direction RA$=168^{\circ}$, Dec$=-7^{\circ}$ (Lineweaver et al. 1996; Hinshaw et al. 2009; Aghanim et al. 2018; Saha et al. 2021). 

However, an unexpected, substantially bigger dipole anisotropy, observed in a large sample of distant radio sources, yielded a peculiar motion $\sim 4$ times (Singal 2011) the CMBR value, though in the same direction as the CMBR dipole. Subsequent confirmations of the peculiar velocity with respect to AGN reference frames being much larger than the CMBR value (Rubart \& Schwarz 2013; Tiwari et al. 2015; Colin et al. 2017; Bengaly, Maartens \& Santos 2018; Singal 2019a,b;  Siewert et al. 2021; Secrest et al. 2021; Singal 2021a,b), 
cast doubts on the CMBR dipole being the ultimate representative of the solar peculiar motion. It has been suggested (Tiwari \& Nusser 2016) that the observed number count excess may mostly be due to structures within $z<0.1$. Also, the spectral index distribution for the NVSS data does not show departures from isotropy over the sky (Ghosh \& Jain 2017). The magnitude of all multipoles, except dipole, determined from a full sky radio galaxy map, seem to be roughly consistent with $\lambda$CDM model (Tiwari \& Aluri 2019). However, an examination of the correlation between the X-ray luminosity and temperature in a homogeneously selected X-ray galaxy cluster sample of 313 objects, over different directions of the sky (Migkas et al. 2020), has shown anisotropies consistent with those seen with other cosmological probes. 
\begin{figure*}
\includegraphics[width=15cm]{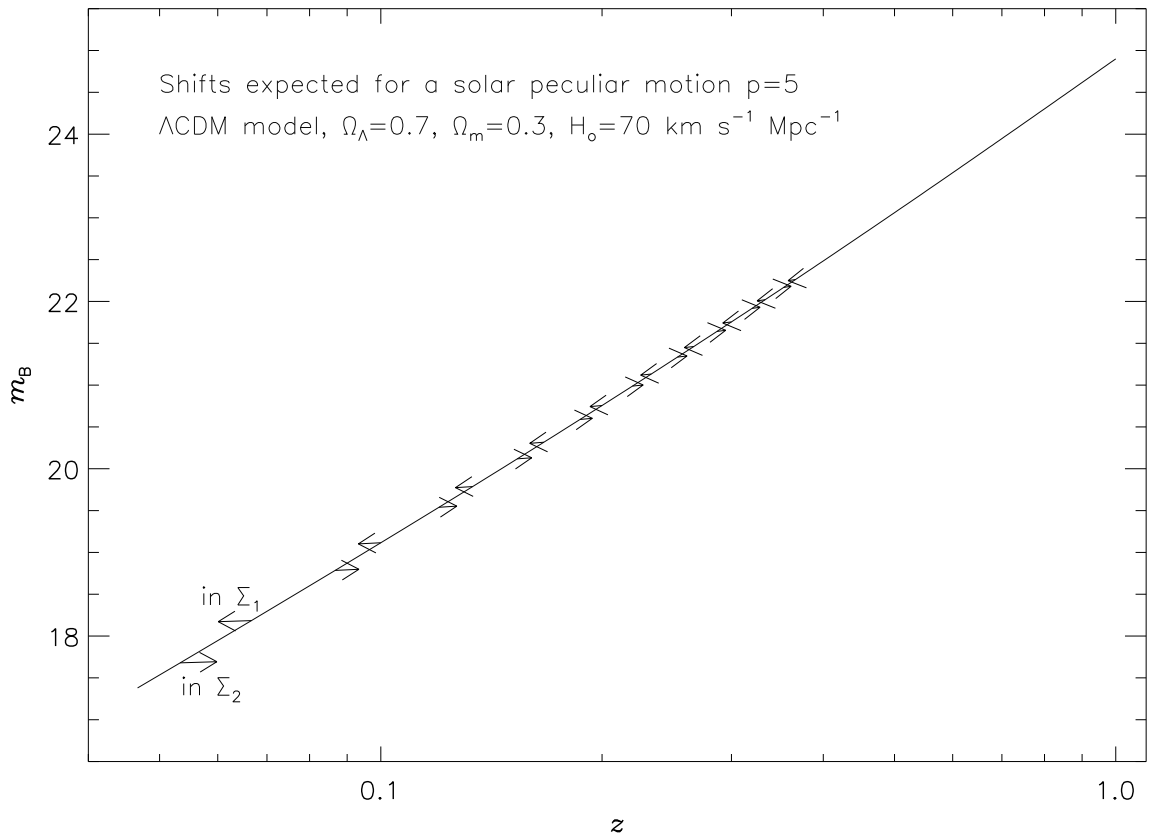}
\caption{ The expected Hubble diagram $m_{\rm o}-z_{\rm o}$, for the SNe Ia, is shown for the standard $\Lambda$CDM homogeneous and isotropic cosmological model. Owing to the peculiar motion of the observer, a source assumedly lying in the forward hemisphere $\Sigma_1$, along the direction of motion, would get displaced to lower redshift, and appear, as a result, at a higher magnitude (fainter) compared to a source that lies in the backward hemisphere $\Sigma_2$ along the anti-pole direction. Arrows represent the displacements that individual sources at particular points on the $m_{\rm B}-z$ diagram would undergo due to observer's  peculiar velocity. For a solar peculiar velocity $p=5$ (quantified in units of the CMBR value of 370 km s$^{-1}$), a supernova Ia at an observed redshift $z=0.06$, along the pole of the peculiar motion in the hemisphere $\Sigma_1$ would be fainter by a magnitude $\Delta m_{\rm B} \approx 0.5$ than a supernova Ia at the same observed redshift along the anti-pole direction in the hemisphere $\Sigma_2$. At higher redshifts, the displacements, represented by the length of the arrows, reduce and the magnitude differences become lesser.}
\end{figure*}

In any case, it might be desirable to determine the peculiar velocity with respect to a different reference frame, and if possible using an independent method, to examine whether the derived peculiar motion is indeed significantly larger than what has been inferred from the CMBR dipole. 
Here we introduce such a novel technique and apply it to determine peculiar motion of the observer, from the magnitude-redshift Hubble diagram for the SNe~Ia, one of the best standard candles known, with an absolute peak blue magnitude $M_{\rm B}=-19.2$ (Richardson et al. 2014;
Efstathiou 2021).
\section{Solar peculiar motion from the Hubble diagram}
If there were no peculiar motion of the Solar system, then the observed ($m_{\rm B}-z$) relation could be employed directly for various cosmological tests. However due to a Solar peculiar motion, there would be alterations in the $m_{\rm B}-z$ plot.  
Due to a peculiar velocity $v$ of the observer, the observed redshift and optical magnitude of an object, lying at an angle $\theta$ with respect to the direction (pole) of the peculiar motion, will get modified as (Davis et al. 2011)
\begin{eqnarray}
\label{eq:80.1}
(1+z)=(1+z_{\rm o})(1-v\cos\theta/c), \\
\label{eq:80.2}
m=m_{\rm o}+5\log(1-v\cos\theta/c), 
\end{eqnarray}
where $z_{\rm o}$, $m_{\rm o}$ are the values as would be measured by a comoving observer, i.e., without any peculiar motion.  Here $v$ is assumed to be non-relativistic ($v\ll c$) since all previous dipole measurements have indicated so. 

For a given peculiar velocity $v$ of the observer, different sources, depending upon their angle $\theta$, will get differently displaced, according to Eqs.~({\ref{eq:80.1}) and ({\ref{eq:80.2}), in the Hubble $m-z$ plot. As the effects on both $m$ and $z$ are proportional to $\cos\theta$, all source with $\cos\theta>0$, and thus lying in the forward hemisphere, say $\Sigma_1$, centred on the pole of peculiar motion, will get displaced in the $m-z$ plot opposite to the sources with $\cos\theta<0$ and thus lying in the backward hemisphere, say $\Sigma_2$, centred on the anti-pole. Accordingly, in the Hubble diagram, there will be a systematic shift between sources belonging to the two hemispheres, $\Sigma_1$ and $\Sigma_2$. This systematic shift provides a measure of the peculiar velocity of the observer, or equivalently that of the solar system.
We use a parameter $p$ to express the amplitude of the peculiar velocity $v$, in units of the CMBR value, so that $v=p \times 370$ km s$^{-1}$, with $p=0$ implying a nil peculiar velocity while $p=1$ implying the CMBR value.

\begin{figure*}
\includegraphics[width=15cm]{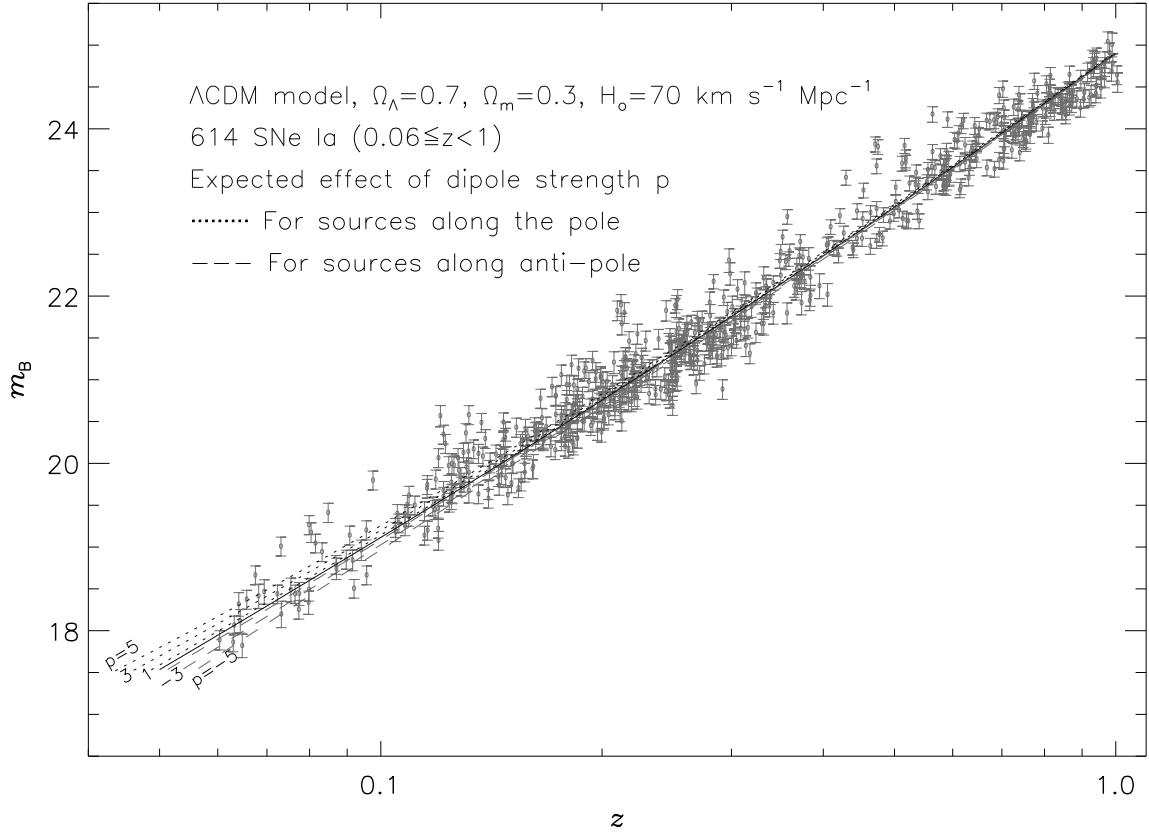}
\caption{The observed Magnitude-redshift ($m_{\rm B}-z$) distribution of SNe~Ia in our sample. The dark continuous line in the middle shows the expected Hubble diagram $m_{\rm o}-z_{\rm o}$, for the SNe Ia, in the standard $\Lambda$CDM homogeneous and isotropic cosmological model. Due to observer's  peculiar velocity, individual sources at any point on this line would get displaced, with the displacement being, to a first order, directly proportional to the amplitude of the peculiar velocity, assumed to be a small non-relativistic value. The family of gray dotted lines at higher $m_{\rm B}$ above the continuous line show the displacements expected for various amplitudes of the peculiar velocity (quantified by $p$, in units of the CMBR value of 370 km s$^{-1}$), the loci of the expected displacements for sources lying in the direction (pole) of the  peculiar velocity at its apex, while the dashed lines below the continuous line show loci of the displacements expected for sources lying in the anti-pole direction, for various $p$ values ($p<0$ in the anti-pole direction).}
\end{figure*}
\begin{figure*}
\includegraphics[width=15cm]{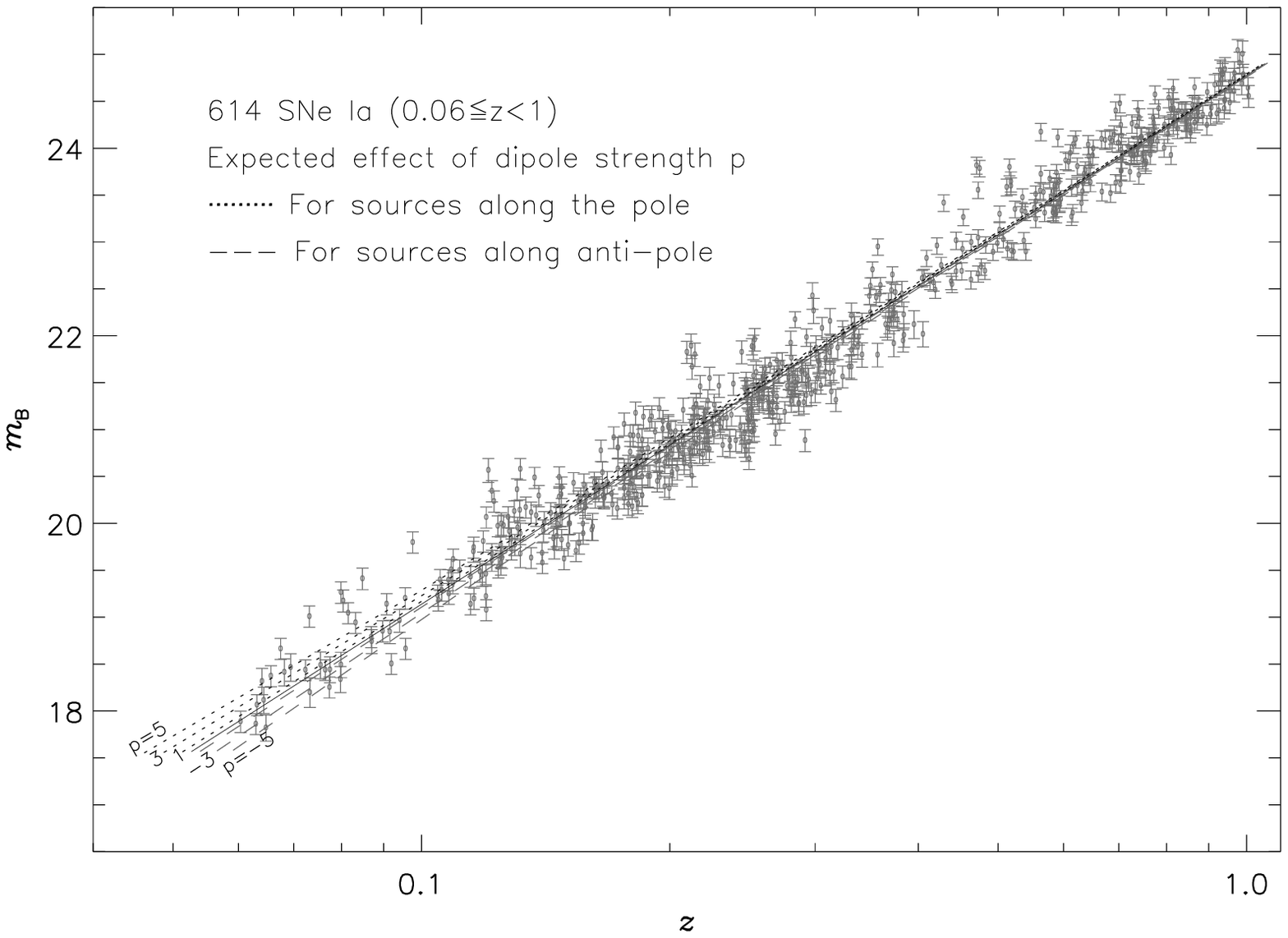}
\caption{The observed Magnitude-redshift ($m_{\rm B}-z$) distribution of SNe~Ia in our sample, ss in Fig.~2.
The dark continuous line in the middle, ostentatiously for $p=0$, shows now a straight line fit with $m_{\rm B} \propto \log z$, to all SNe~Ia in our sample, without assuming any particular $\Lambda$CDM cosmology. Due to observer's  peculiar velocity, individual sources at any point on this line would get displaced, with the displacement being, to a first order, directly proportional to the amplitude of the peculiar velocity, assumed to be a small non-relativistic value. The family of gray dotted lines at higher $m_{\rm B}$ above the continuous line show the displacements expected for different amplitudes of the peculiar velocity (quantified by $p$, in units of the CMBR value of 370 km s$^{-1}$), the loci of the expected displacements for sources lying in the direction (pole) of the  peculiar velocity at its apex, while the dashed lines below the continuous line show loci of the displacements expected for sources lying in the anti-pole direction, for different $p$ values ($p<0$ in the anti-pole direction).}
\end{figure*}

In the absence of a peculiar motion of the observer, the magnitude, $m_{\rm o}$, of a source that one would observe at a redshift $z_{\rm o}$, could be computed for any given cosmological model from the formula (Efstathiou 2021)
\begin{eqnarray}
\label{eq:4}
m_{\rm o} = 25 + 5 \log D_{\rm L}(z_{\rm o}) + M
\end{eqnarray}
where $D_{\rm L}(z_{\rm o})$ is the luminosity distance (Mpc) at redshift $z_{\rm o}$ in that cosmological model and $M$ is the absolute magnitude of the source. 

We adopt here the $\Lambda$CDM cosmology, as an example, to calculate the luminosity distance $D_{\rm L}$. 
In the standard $\Lambda$CDM homogeneous and isotropic cosmological model, the space is flat ($k=0$) with the density parameter $\Omega=\Omega_{\rm m}+\Omega_{\rm r}+\Omega_\Lambda=1$ with $\Omega_{\rm m}$,  $\Omega_{\rm r}$, $\Omega_\Lambda$ being the matter density, radiation density and vacuum energy (dark energy) density parameters respectively (Aghanim et al. 2020; Hobson, Efstathiou \& Lasenby 2006; Weinberg 2008)
The luminosity distance $D_{\rm L}$ of a source, in a matter-dominated universe ($\Omega_{\rm r}=0$), can then be evaluated in terms of the cosmological redshift $z_{\rm o}$ by a numerical integration (Hobson et al. 2006; Weinberg 2008)
\begin{eqnarray}
\label{eq:5}
D_{\rm L}(z_{\rm o})_= \frac{c(1+z_{\rm o})}{H_{\rm o}}\int^{1+z_{\rm o}}_{1}\frac{{\rm d}z}{\left[\Omega_\Lambda+\Omega_{\rm m}z^3\right]^{1/2}}.
\end{eqnarray}
We use a Hubble constant $H_{\rm o}=70\,$km~s$^{-1}$\,Mpc$^{-1}$, the matter energy density $\Omega_{\rm m}=0.3$, and 
the vacuum energy (dark energy!) density $\Omega_{\Lambda}=0.7$ (Hinshaw, et al. 2009; Aghanim et al. 2020)

From Eq.~({\ref{eq:80.2}) one would expect the sources to be brighter in the forward hemisphere, $\Sigma_1$. However, because of the shift in redshift due to the peculiar motion (Eq.~({\ref{eq:80.1})), the net effect at an observed redshift is the opposite. Figure~1 shows schematically the  shifts expected at some representative points in the $m_{\rm B}-z$ Hubble diagram for the SNe Ia, one of the best standard candles known with an absolute peak blue magnitude $M_{\rm B}=-19.2$ (Richardson et al. 2014; Efstathiou 2021),
in the standard $\Lambda$CDM homogeneous and isotropic cosmological model. Arrows show the displacement that individual sources at particular points on the $m_{\rm B}-z$ diagram would undergo, for a peculiar velocity, $p=5$ (quantified in units of the CMBR value of 370 km s$^{-1}$). As can be seen from Fig.~1, a source assumedly lying in the hemisphere $\Sigma_1$ would get displaced to a lower redshift, and as a result appear at a higher magnitude (fainter) compared to an equivalent source that lies in the hemisphere $\Sigma_2$. For instance, as can be seen from Fig.~1, for a solar peculiar velocity p=5, a supernova Ia at a redshift $z=0.06$, seen along the pole of the peculiar motion would appear fainter by a magnitude $\Delta m_{\rm B} \approx 0.5$ than a supernova Ia at the same observed redshift in the anti-pole direction. As we go to higher redshifts, the displacements, represented by the length of the arrows in Fig.~1, reduce and the magnitude differences, $\Delta m_{\rm B}$, steadily decrease. However, all the sources that were originally on the line representing the $m_{\rm o}-z_{\rm o}$ relation, and belonged to the $\Sigma_1$ hemisphere, will  shift above the line and would thus appear fainter, while the sources  on this line in $\Sigma_2$ hemisphere would move below the line and appear brighter. Thus at all observed redshifts, SNe Ia lying in the hemisphere $\Sigma_1$ will be systematically fainter than those observed at similar redshifts in the hemisphere $\Sigma_2$, though the difference will be more pronounced at lower redshifts.

\begin{figure*}
\includegraphics[width=15cm]{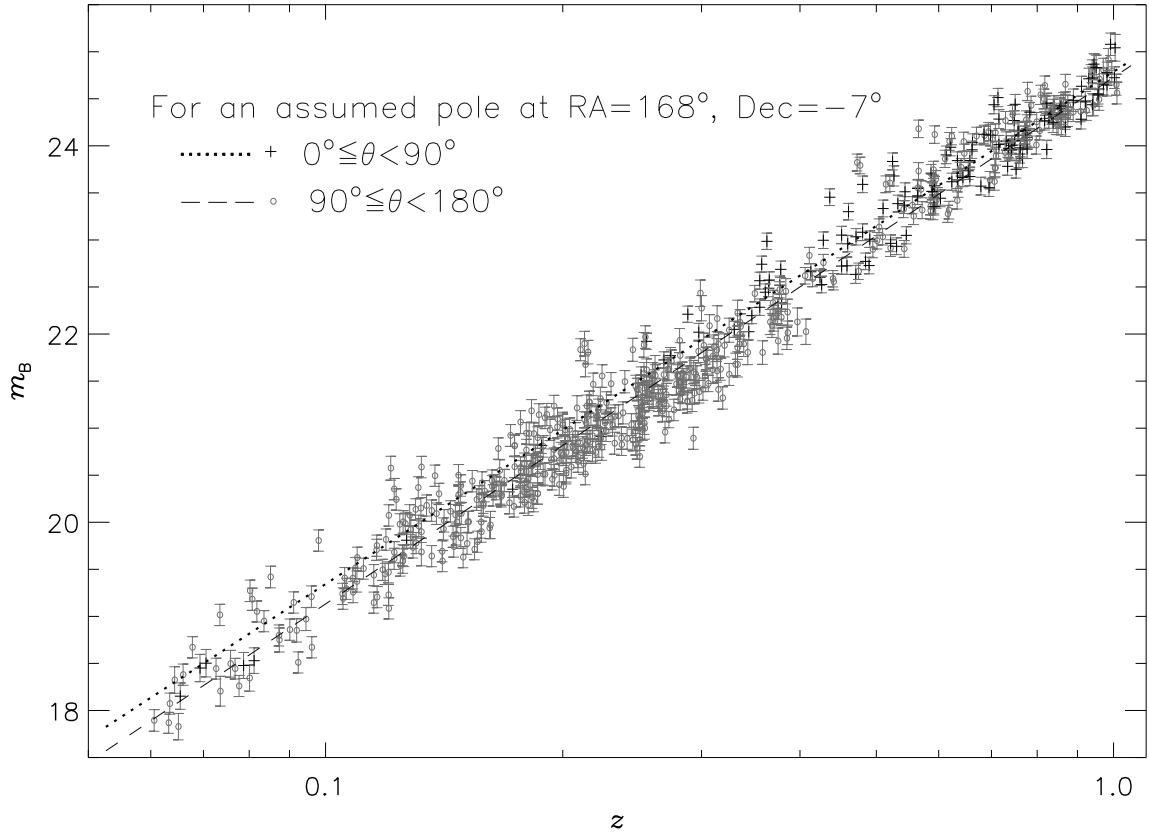}
\caption{The observed Magnitude-redshift ($m_{\rm B}-z$) distribution of SNe~Ia in the sky.
The dotted line shows the straight line fit to SNe~Ia, denoted by plus (+) symbols, in the $\Sigma_1$ hemisphere centred on the CMBR pole, while the dashed line shows the straight line fit to SNe~Ia, denoted by circles (o), in the $\Sigma_2$ hemisphere, centred on the CMBR anti-pole. The two lines show from each other displacements in magnitude, especially at low redshifts, say at $z\sim 0.06$. The displacement $\Delta m_{\rm B}$, to a first order, is proportional to (the component of) the observer's peculiar velocity in the CMBR dipole direction, and thus is a measure of $p \cos\psi$, with $\psi$ being the angle of the  CMBR dipole from the true dipole direction.}
\end{figure*}
\begin{figure*}
\includegraphics[width=15cm]{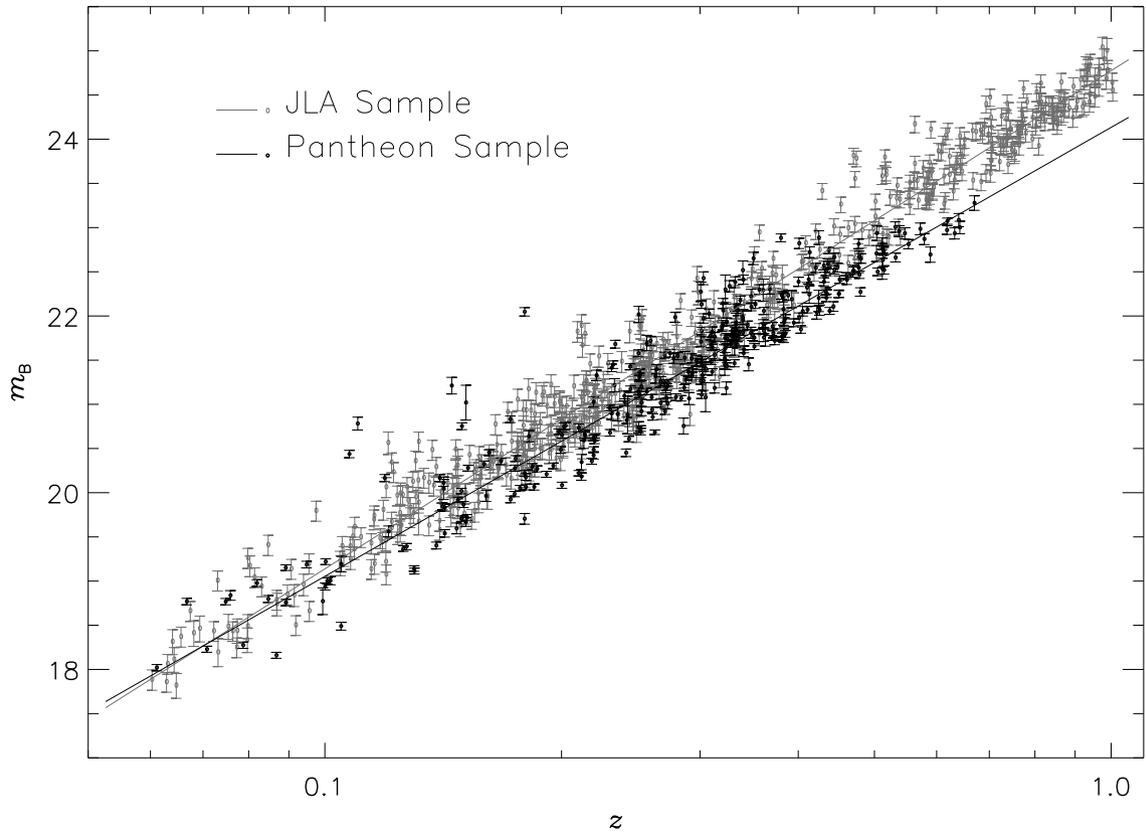}
\caption{A comparison of the Magnitude-redshift ($m_{\rm B}-z$) distribution of SNe~Ia in the JLA (fainter gray points) and Pantheon samples (darker points). Although the two samples seem to match at low redshifts ($z\sim 0.06$), at higher redshifts they progressively depart in magnitude ($m_{\rm B}$), so much that at $z\sim 0.6$, the Pantheon sample is systematically brighter than the JLA sample by  $\Delta m_{\rm B}\approx -0.5$.
}
\end{figure*}
\begin{figure*}
\includegraphics[width=\linewidth]{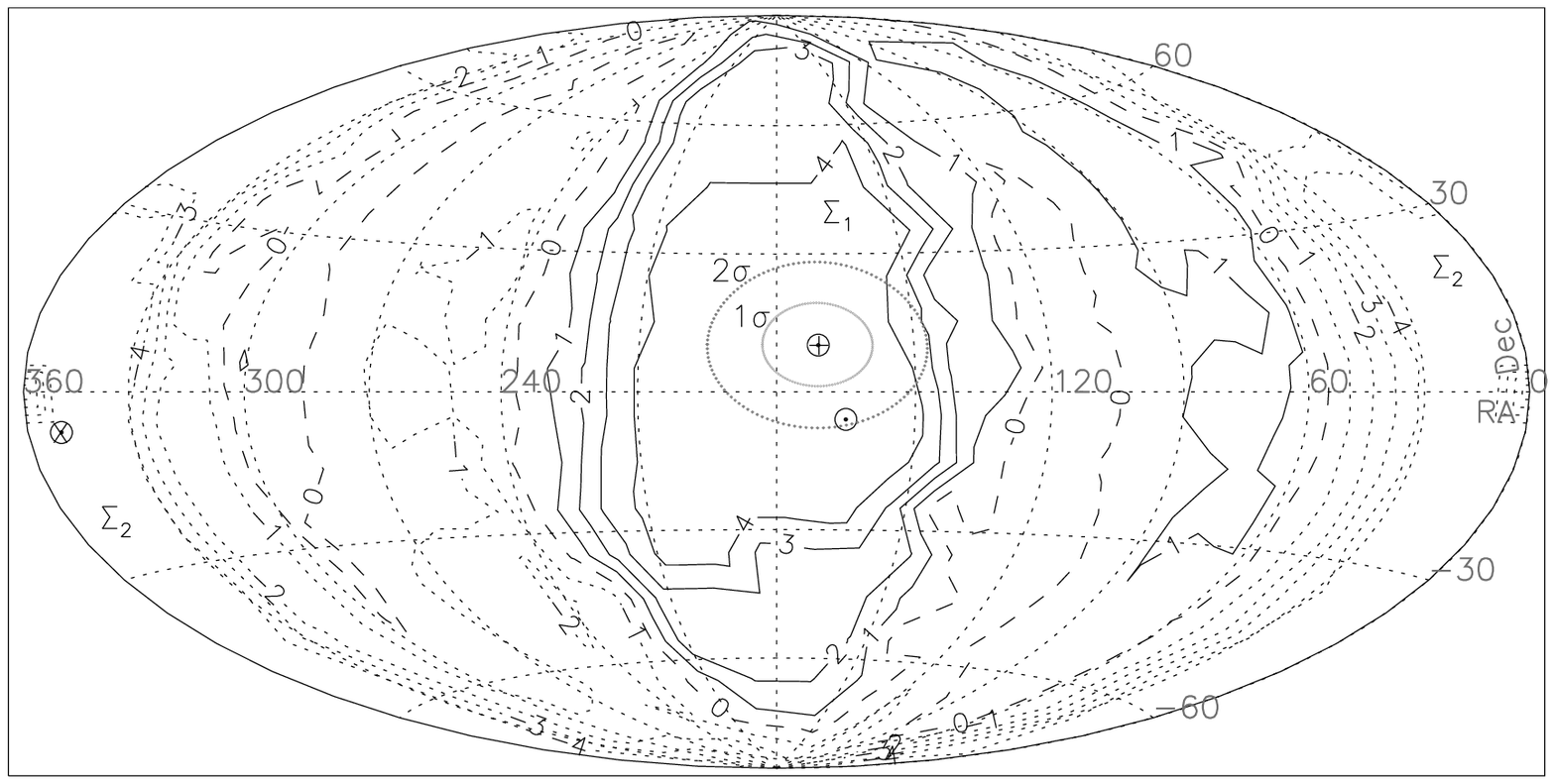}
\caption{A contour map of the dipole amplitudes, in the Hammer--Aitoff equal-area projection on the sky. The~horizontal and vertical axes denote RA,  from $0^\circ$ to $360^\circ$, and declination (Dec), from $-90^\circ$ to $90^\circ$. The~true pole direction is expected to be closer to the higher contour values, indicated by continuous lines, while the true antipole should lie closer to the lower contour values, indicated by dotted lines. Dashed lines represent the zero contour values.
The symbol $\oplus$ indicates the best-fit pole position for the SNe Ia sample, derived using our 3-d cosfit routine (see the text), while the symbol $\otimes$ indicates the corresponding antipole position. Two gray-color error ellipses  around the best-fit position $\oplus$ represent the $1\sigma$ ($68.3\%$) and $2\sigma$ ($95.5\%$) confidence limits. The symbol $\odot$ indicates the CMBR pole position, which lies within $2\sigma$ of the derived pole position $\oplus$ for our SNe Ia sample.
} 
\end{figure*}

This thus provides us a novel technique, which we could apply to determine the peculiar motion of the solar system from the magnitude-redshift Hubble diagram for SNe~Ia, which are one of the best standard candles known, with a very tight  $m_{\rm B}-z$ relation, where $m_{\rm B}$ denotes the observed peak blue magnitude and $z$ the measured redshift of each SN~Ia. As we shall see, this technique, unlike other methods employed to estimate our peculiar motion, e.g. number counts or sky brightness from AGN surveys comprising $\stackrel{>}{_{\sim}} 10^5$ sources (Singal 2011;  Gibelyou \& Huterer 2012; Rubart \& Schwarz 2013; Tiwari et al. 2015; Colin et al. 2017; Bengaly et al. 2018; Singal 2019a,b; Siewert et al. 2021; Secrest et al. 2021; Singal 2021a,b), 
can yield statistically significant results from much smaller number ($\stackrel{<}{_{\sim}} 10^3$)  of SNe. Moreover, the completeness of the survey, an absolute requirement in other methods where all sources above a certain observed flux density limit are part of the sample, is not a prerequisite. Nor is the full sky-coverage essential; a piece-wise coverage of the sky in different directions could suffice. In fact, it could as such be applied to a combination of data from a heterogeneous set of various sub-samples; all that is required is that no systematic errors have observationally entered in the redshift and magnitude estimates of individual sources lying in different directions in sky.  

\section{Our sample of SNe Ia and the procedure deployed for computing the peculiar motion}
For our purpose, we have selected a restricted sub-sample of the JLA sample (Betoule et al. 2014)
which contained 740 spectroscopically confirmed SNe Ia, spanning a  redshift range 0.01-1.3. However, we have restricted for our purpose the lower limit to 0.06 in order to keep the effect of local bulk flows to a minimum (Colin et al. 2019).
Further, there are only 8 SNe~Ia in the JLA sample with $z>1$, which we have excluded leaving us with a total of 614 SNe~Ia in our sample. The `corrections' already applied to the redshifts and magnitudes based on the traditionally adopted peculiar velocity of the solar system, $370$ km s$^{-1}$ along RA$=168^{\circ}$, Dec$=-7^{\circ}$, as derived from the CMBR dipole, have been reverted to get back the observed heliocentric redshifts and magnitudes, utilizing Eqs.~({\ref{eq:80.1}) and ({\ref{eq:80.2}).

Figure~2 shows the peak blue magnitude versus redshift ($m_{\rm B}-z$) plot for our sample of SNe~Ia, where we have also drawn the expected Hubble diagram,  $m_{\rm o}-z_{\rm o}$, for SNe Ia, adopting an absolute peak blue magnitude $M_{\rm B}=-19.2$ (Richardson et al. 2014;
Efstathiou 2021), 
in the standard $\Lambda$CDM homogeneous and isotropic cosmological model. Figure~2 demonstrates how the expected plots will shift in the $m_{\rm B}-z$ plane for sources in the two hemispheres for various $p$ values, and as can be seen from the figure, any shifts in $m_{\rm B}-z$ plots at large redshifts (as $z\rightarrow 1$) are relatively insignificant because combined effects of the Eqs.~(\ref{eq:80.1}) and (\ref{eq:80.2}) at larger redshifts turn out to be rather small, as was seen by the lengths of arrows representing displacements in Fig.1.

In Fig.~2, the continuous line, which is the Hubble diagram for SNe Ia in the standard $\Lambda$CDM homogeneous and isotropic cosmological model, is thus for $p=0$, that is when there were no Solar peculiar motion. However, due to our  peculiar velocity  
individual sources at any point on this line in the $m_{\rm B}-z$ diagram would get displaced, with the displacement being, to a first order, directly proportional to the amplitude of the peculiar velocity, assumed to be a small non-relativistic value. 
The set of grey dotted lines above (at $m_{\rm B}$ values higher than) the $p=0$ line show, for an increasing $p$, the 
loci of expected displacements for sources lying in the hemisphere $\Sigma_1$, while the grey broken lines below the continuous line show loci of the displacements expected for sources in $\Sigma_2$ ($p<0$ in this direction).
Thus a difference in magnitude $\Delta m_{\rm B}$ between sources lying in  $\Sigma_1$ and $\Sigma_2$ at some given low enough redshift, say $z=0.06$, yields value of $p$ for the peculiar motion. It should be noted that in Fig.~2, plots depicted for various $p$ values are for sources along the pole or anti-pole direction. For an even distribution of sources along various directions within each hemisphere, the net displacement will on the average be half of that shown in Fig.~2 for each $p$ value.

For any $z_0$ and $m_0$ values, what matter are the $\Delta z$ and $\Delta m$ values, which  from Eqs.~(\ref{eq:80.1}) and (\ref{eq:80.2}), depend only on the component of the peculiar velocity $v \cos \theta$ and the slope of the Hubble diagram. Therefore, as long as the slope of the $m_{\rm B}-z$ plot among various cosmological models does not vary significantly, especially at low $z$ values where different cosmological models do not differ much, only  $v \cos \theta$ would  mainly be the quantity in deciding $\Delta m$ between the sources in the two hemispheres. In Fig.~3, the dark continuous line, which is an  empirical straight line fit to the observed $m_{\rm B} \propto \log z$ data, can be taken to be for $p=0$, since there may be an overlap of sources distributed evenly in both $\Sigma_1$ and $\Sigma_2$ hemispheres. Like in Fig.~2, here too the set of grey dotted lines above (at $m_{\rm B}$ values higher than) the $p=0$ line show, for an increasing $p$, the 
loci of expected displacements for sources lying in the hemisphere $\Sigma_1$, while the grey broken lines below the continuous line show loci of the displacements expected for sources in $\Sigma_2$ ($p<0$ in this direction).
Thus a difference in magnitude $\Delta m_{\rm B}$ between sources lying in  $\Sigma_1$ and $\Sigma_2$ at some given low enough redshift, say $z=0.06$, yields value of $p$ for the peculiar motion. It should be noted that in Fig.~3, plots depicted for various $p$ values are for sources along the pole or anti-pole direction. 

A comparison of Figs. 2 and 3 shows the similarity in the two diagrams, as far as the differences between the two hemispheres is concerned.  Therefore we  can use only a simple empirical relation, a straight line,  between $m_{\rm B}$ and $\log z$, and which, as seen from Fig.~3, yields quite a good fit. Even otherwise, a higher degree polynomial fit than a straight line may not be essential for our purpose, as the displacements in the $m_{\rm B}-z$ plot, as evident from Fig.~3, are predominantly at lower redshifts, where differences between various cosmological models are not significant. 
\begin{figure*}
\includegraphics[width=\linewidth]{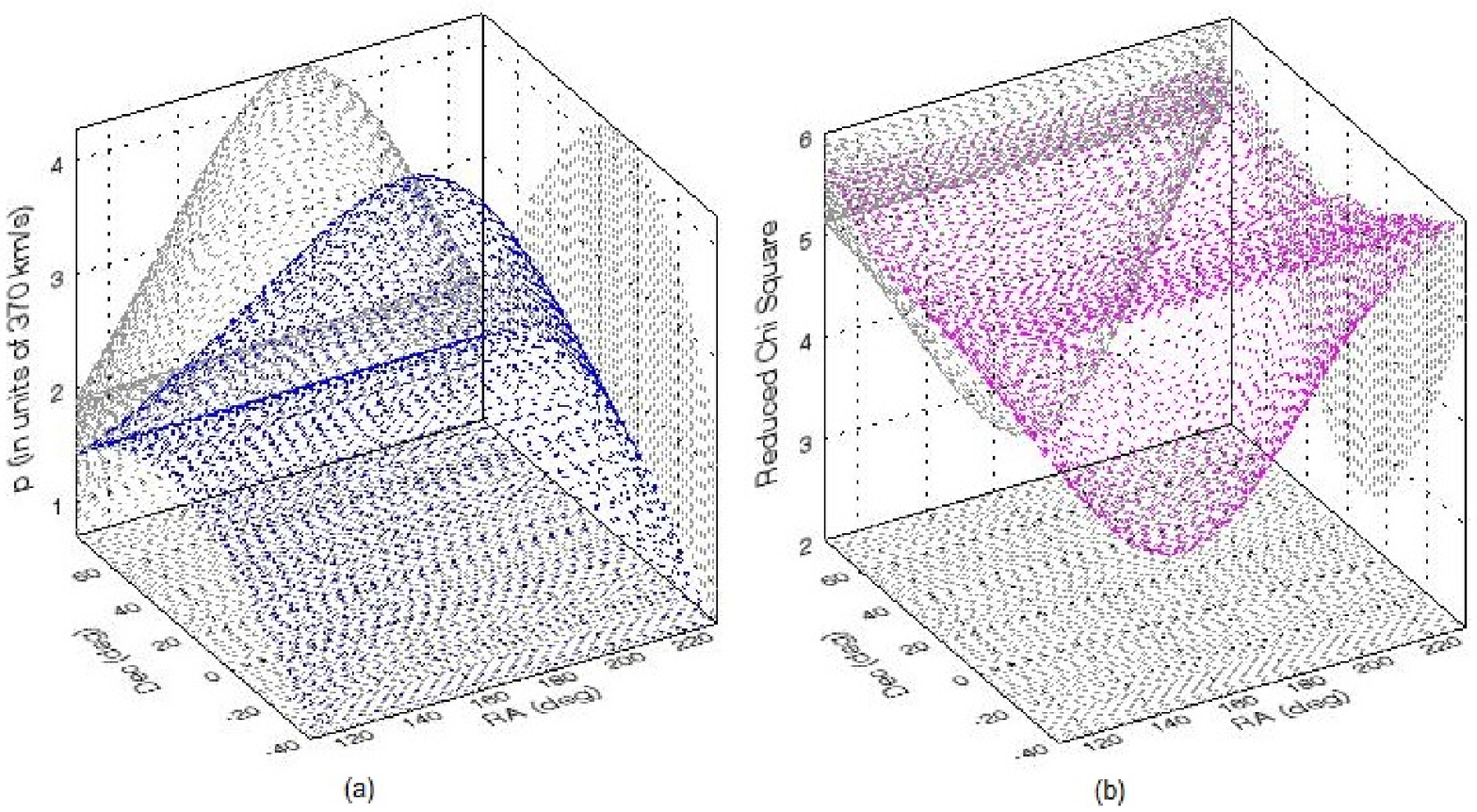}
\caption{A plot of 3-d  cosfit made to the estimated $p$ values for various trial dipole directions across the sky, showing (a) a unique peak (blue colour) along RA$=175^{\circ}$, and Dec$=8^{\circ}$, which unambiguously indicates the optimum direction of the dipole (b) reduced chi-square ($\chi^2_\nu$) (violet colour), having a minimum value of 2.5 (somewhat higher than the ideal value of unity) at RA$=171^{\circ}$, and Dec$=12^{\circ}$. The horizontal axes denote RA and Dec in degrees. 
The positions (RA and Dec) of the extrema are determined more easily from the 2-d projections, shown in light grey. Thence we infer the direction of the observer's peculiar velocity as RA$=173^{\circ}$, and Dec$=10^{\circ}$, within $\sim 2\sigma$ of the CMBR dipole direction, however the amplitude appears to be a factor $\sim 4$ higher than the CMBR value.} 
\end{figure*}

To begin with, we first assume that the peculiar velocity is along the CMBR dipole, RA$=168^{\circ}$, Dec$=-7^{\circ}$. 
Then using the great circle at $90^\circ$ from this pole direction, we divide the sky in two equal hemispheres, $\Sigma_1$ and $\Sigma_2$, with $\Sigma_1$ containing the above pole, and $\Sigma_2$ containing the anti-pole.  
Figure~4 shows the actual displacement that occurs in the $m_{\rm B}-z$ diagram between sources in two opposite hemispheres, taking the peculiar motion to be along the CMBR dipole. The dotted line above and the dashed line below depict a fit to the actual  $m_{\rm B}-z$ plots for sources, separately in the $\Sigma_1$ and $\Sigma_2$ hemispheres. From a comparison with the loci in Fig.~3 of the expected displacements for sources lying in the two hemispheres, the observed displacement between the dotted and dashed lines in Fig.~4 suggests $p\stackrel{>}{_{\sim}} 4$, instead of $p=1$, expected for the CMBR dipole. The inferred $p$ value, at least to a first order, will be proportional to $\cos \psi$, the projection of the assumed dipole, CMBR here, on the actual dipole, if the latter is along a different direction.
 It is evident that the peculiar velocity is certainly not concordant with the CMBR value, which should have given $p\approx 1$.

In our sample there are only 128 SNe in the hemisphere $\Sigma_1$ while 486 SNe belong to $\Sigma_2$, in fact, there is a particular deficiency in $\Sigma_1$ of sources in the $z$ range 0.08 to 0.3. 
This comes from the sky coverage in various sub-samples of the JLA sample. 
A somewhat even split of sources amongst $\Sigma_1$ and  $\Sigma_2$ would minimize statistical uncertainties in the mutual displacement estimates. In order to have additional sources  in the $\Sigma_1$ hemisphere,  
we examined the Pantheon sample of 300 plus SNe 
(Jones et al. 2018; Scolnic et al. 2018) 
for inclusion in our study. However, it has been reported that the Pantheon sample may have significant discrepancy in redshift values as compared to the JLA sample data (Rameez 2019).
To investigate suitability of the Pantheon samples for our purpose, we made the $m_{\rm B}-z$ plot of SNe~Ia belonging to both JLA and Pantheon samples separately in the same diagram. Figure~5 shows the $m_{\rm B}-z$ plots and straight line fits for both samples. In order to distinguish the two plots, we have used light gray points for the JLA data and also shown the straight line fit to the  $m_{\rm B}-\log z$ plot by a light colour line as compared to that for the fit to the Pantheon data. It does seem that there is a systematic difference in the two data. Although the two samples seem to match at low redshifts ($z\sim 0.06$), at higher redshifts they progressively depart in magnitude ($m_{\rm B}$), so much that at $z\sim 0.6$, the Pantheon sample is systematically brighter than the JLA sample by  $\Delta m_{\rm B}\approx -0.5$.
It should be noted that the data in both cases has been reduced to the Heliocentric system by reverting any corrections that had been made to  $m_{\rm B}$ as well as $z$ for the assumed peculiar motion with respect to the CMBR. It is not clear what Hidden or as yet not-understood systematics in either of the two samples are causing differences between JLA and Pantheon data samples. For the present purpose we shall confine ourselves to the JLA sample alone, as this has been employed more often in past for cosmological studies.
\section{DIRECTION OF THE PECULIAR MOTION}
In Fig.~4, we had assumed the peculiar motion to be along the CMBR dipole, however, the actual direction of the peculiar motion might be different. To get a handle on the true direction of the SNe~Ia dipole, without a bias toward any particular direction, including that toward the CMBR dipole, we employ `the brute force method' 
(Singal 2019b), 
We divide the sky into pixels of $2^\circ \times 2^\circ$, creating a grid of $10360$ cells covering the whole sky area of $4\pi$ sr ($=41253$ square degrees), with minimal overlaps. 
Then one by one, taking the trial pole direction to be the centre of each of these 10360 pixels, and accordingly dividing our sample of 614 SNe~Ia into two hemispheres, $\Sigma_1$ and $\Sigma_2$ with respect to that trial pole direction, we compute the dipole amplitudes ($p$), along with standard errors, from the best fits to the $m_{\rm B}-z$ data. Thus for each of $10360$ pixels, we have RA, Dec, and a peculiar velocity value $p$. However, this $p$ value represents only a projection of the true peculiar velocity along that specific RA and Dec. Therefore, we can expect a peak along the real dipole direction, along with a $\cos\psi$ dependence in the $p$ values, determined for various grid points around it.

The dipole amplitude distribution across the sky, as obtained by us for all $10360$ cells, is shown in a contour map (Fig.6). 
The~location of the peak value for the dipole amplitude, in principle, should yield the true direction of the dipole. A broad plateau showing maxima in $p$, towards certain directions near the CMBR dipole direction is clearly seen, however, from that it is not possible to zero down on a single unique peak for the true dipole direction. Nevertheless, we can refine the procedure for  determining the pole direction by making use of the expected $\cos\psi$ dependence of $p$ for grid points at polar angle $\psi$ from the true pole. Thus for each of the $n=10360$ sky positions, we made a 3-d cosfit to the $p$ values of surrounding $n-1$ around it, and determined the pixel with the highest value, which should yield the optimum direction for the peculiar motion.

For this we one by one chose a grid point, say $i$th, out of $n$ grid points, and determined angle $\psi_{\rm ij}$ between the chosen $i$th pixel and each one of the remaining ($j=1$ to $j=n-1$) grid points, and computed the amplitude $a(i)$ by minimizing $\chi^2(i)$, given by (see e.g., Bevington \& Robinson 2003)
\begin{eqnarray}
\label{eq:80.6}
\chi^2(i)= \Sigma^{n-1}_{j=1}\Big[\frac{p_{\rm j}-a(i)\cos \psi_{\rm ij}}{\sigma_{\rm j}}\Big]^{2}\,,
\end{eqnarray}
where $\sigma_{\rm j}$ is the uncertainty in the estimate of $p_{\rm j}$ for $j$th pixel. The optimum direction for the peculiar motion got identified with the pixel $i$ that yielded the highest amplitude $a(i)$.

We also compared reduced $\chi^2$ values for each of the $n$ pixels, 
\begin{eqnarray}
\label{eq:80.7}
\chi^2_{\nu}(i)= \frac{1}{n-2}\Sigma^{n-1}_{j=1}\Big[\frac{p_{\rm j}-a(i)\cos \psi_{\rm ij}}{\sigma_{\rm j}}\Big]^{2}\,,
\end{eqnarray}
The minima of the $n$ reduced $\chi^2$ values should be for the pixel $i$ that lies close to the true direction of the peculiar motion and for realistic $\sigma_{\rm j}$ estimates, the value of the minimum of reduced $\chi^2$ should be around unity. 

The process converged to show an unambiguous peak for the optimum dipole at RA$=175^{\circ}$, Dec$=8^{\circ}$ (Fig.~7a). Also the reduced $\chi^2$ showed a unique minimum at RA$=171^{\circ}$, Dec$=12^{\circ}$ (Fig.~7b), in quite close vicinity of the optimum dipole. 
A grid size of $1{^\circ} \times 1{^\circ}$, with more than $41000$ cells, 
or a still finer grid size of $0\!\!\stackrel{^\circ}{_.}\!\!5 \times 0\!\!\stackrel{^\circ}{_.}\!\!5$ made no perceptible difference in our results. From these, we take our peculiar velocity to be average of these two values, RA$=173^{\circ}$, Dec$=10^{\circ}$.
\begin{figure*}
\includegraphics[width=\linewidth]{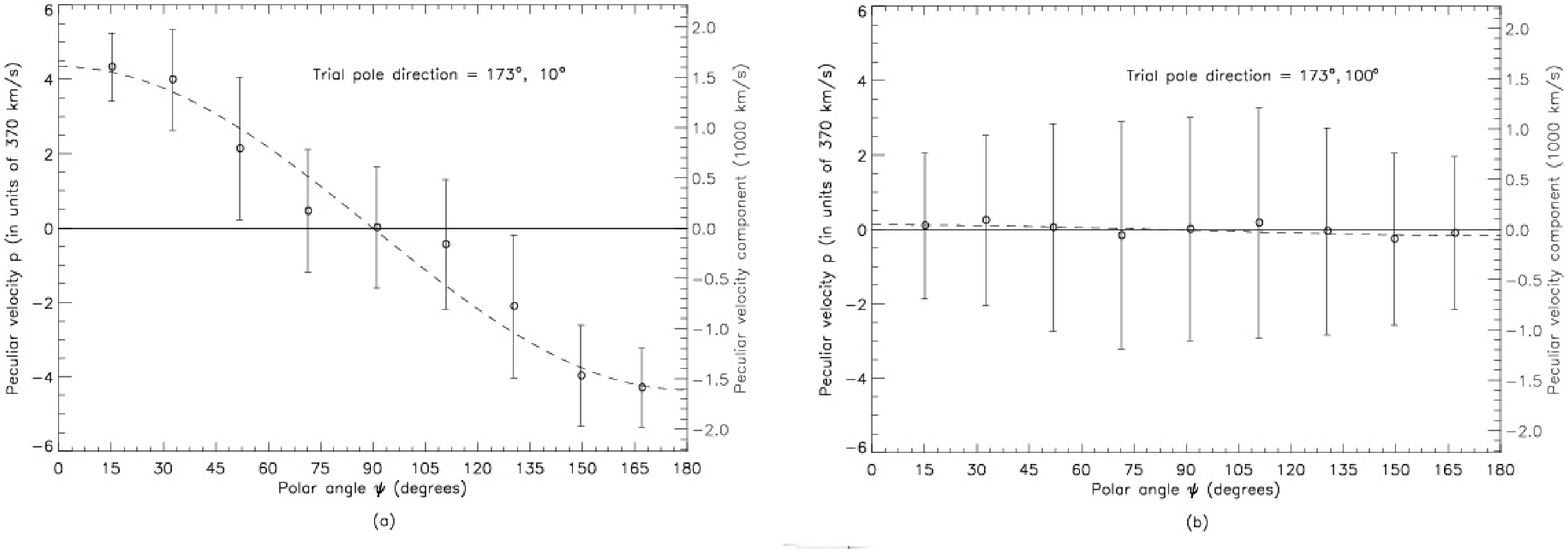}
\caption{Variation of the peculiar velocity component $p$ (in units of CMBR value 370 km s$^{-1}$), computed for various polar angles (a) with respect to the derived best-fit direction, RA$=173^{\circ}$, Dec$=10^{\circ}$ (b) with respect to a pole tried at RA$=173^{\circ}$, Dec$=100^{\circ}$, which is at $90^{\circ}$ from the best-fit pole direction. The corresponding peculiar velocity values of the solar system in units of $10^3$ km s$^{-1}$ are shown on the right hand vertical scales.
Plotted circles (o) with error bars in each case show values for bin averages of the peculiar velocity components, obtained for various $20^{\circ}$ wide slices of the sky in polar angle. 
 The dashed line shows a least square fit of $\cos \psi$ to the bin average values.}
\end{figure*}

In order to make sure that the peculiar velocity values computed for neighbouring pixels around the best-fit trial pole do have a $\cos\psi$ dependence, we divided the sky into bins of $20^\circ$ width in polar angle about our determined best-fit pole position, RA$=173^{\circ}$, Dec$=10^{\circ}$, and computed the bin averages of the peculiar velocity $p$, obtained for various $20^{\circ}$ wide slices of the sky. 
A least square fit of $\cos \psi$ to the bin average values, as shown in Fig.~8a, justifies our assumption that the computed $p$ values for various pixels at polar angles ($\psi$) do follow  a 
systematic $\cos\psi$ dependence. A similar exercise with respect to a trial pole at a pixel position, RA$=173^{\circ}$, Dec$=100^{\circ}$, which is at $90^{\circ}$ from the best-fit pole direction, does not show any such systematic $\cos\psi$ dependence Fig.~8b.

To validate our cosfit procedure, we made Monte Carlo simulations. Since our sample is far from
isotropically distributed on the celestial sphere, in our simulations we preserved the sky  positions (RA and Dec) of various SNe in our sample but reallotted the observed redshift and magnitude ($m_{\rm B},z$) pairs randomly amongst our sources in the sample. Then a mock dipole of a randomly chosen value was superimposed to calculate $z$ and $m_{\rm B}$ for each source according to Eqs.~({\ref{eq:80.1}) and ({\ref{eq:80.2}). On this mock catalogue of SNe~Ia, our procedure was applied to recover the dipole and compared with the input mock dipole in that simulation. This not only tested our procedure and the routine but it also provided us a method for making estimates of the uncertainties of the recovered dipole amplitude and direction. This way we made 2500 independent Monte Carlo simulations in five sets of 500 each so as to be sure that results from simulations are consistent across different sets of 500 simulations. In one of the sets, a mock dipole equal to the CMBR value, i.e. p=1, was attempted which, as expected, gave very uncertain results since errors became comparable to the signal adopted. In the last set of 500 simulations, we used the mock dipole to be the same as our derived dipole which was superposed on our simulated mock catalogue of SNe Ia, with original sky positions maintained but with randomly redistributed $m_{\rm B},z$ pairs, to get realistic estimates of the uncertainties. 

This way we obtained our peculiar velocity to be along RA$=173^{\circ}\pm 12^{\circ}$, Dec$=10^{\circ}\pm 9^{\circ}$, with $p=4.3\pm 1.3$,  a $3.3 \sigma$ result.  The corresponding error ellipses at $1\sigma$ and $2\sigma$ levels are shown in Fig.~6, indicating respectively the $68.3\%$ and $95.5\%$ confidence limits. Also plotted is the CMBR pole position (RA$=168^{\circ}$, Dec$=-7^{\circ}$), which lies at the highest contour level in Fig.~6, and our best-fit position of the SNe Ia dipole is within $2\sigma$ of the CMBR dipole position. 

\section{Results and Discussion}
Following a simple kinematic approach, without ascribing to any particular cosmological model, we have attempted to extract out the purely special relativistic or even non-relativistic Doppler effects of the observer's (Solar system's) peculiar motion, by making an appropriate (first degree) polynomial fit to the Hubble plot.
In fig.~9 are shown the dotted and dashed lines, fits to sources in $\Sigma_1$ and  $\Sigma_2$ hemispheres respectively, which after the corrections for the above dipole is applied, almost coincide, especially at lower redshifts, where a large mutual displacement had otherwise appeared between sources belonging to $\Sigma_1$ and  $\Sigma_2$, because of observer's peculiar velocity. To demonstrate it more explicitly, in Fig.~10 we have plotted the expected differences $\Delta m_{\rm z}$ between average magnitudes of sources from the two hemispheres, $\Sigma_1$ and $\Sigma_2$, plotted as dotted lines, for mock peculiar velocity values ($p=0$ to 7) of the observer along RA=173$^\circ$, Dec=10$^\circ$. 
The unbroken line shows a fit to the actual observed difference at different redshifts in the average magnitudes of sources from the two hemispheres, again for the peculiar velocity direction along RA=173$^\circ$, Dec=10$^\circ$, while the dashed lines above and below the unbroken line represent the $1 \sigma$ uncertainties in the fit. 

The SNe~Ia dipole direction lies within $\stackrel{<}{_{\sim}} 2\sigma$ of the CMBR dipole direction, RA$=168^{\circ}$, Dec$=-7^{\circ}$. However, the amplitude of the velocity, $1.6\pm 0.5\times 10^3$ km s$^{-1}$, is $\stackrel{>}{_{\sim}} 4$ times the CMBR value. Our derived motion though seems to be in excellent agreement with the value
 derived (Singal 2011; Rubart \& Schwarz 2013; Tiwari et al. 2014; Colin et al. 2017) from the NRAO VLA Sky Survey (Condon  et al. 1998) data 
  as well as that derived recently (Singal 2021a,b)
  from the number counts of mid infra red AGNs (Secrest, et al. 2015),
but is smaller than that (Bengaly et al. 2018; Singal 2019a; Siewert et al. 2021) derived from the TIFR GMRT Sky Survey data (Swarup et al. 1991; Intema et al. 2017) or the DR12Q data (Singal 2019b) from the Sloan Digital Sky Survey III (P\^aris  et al. 2017). 
\begin{figure*}
\includegraphics[width=15cm]{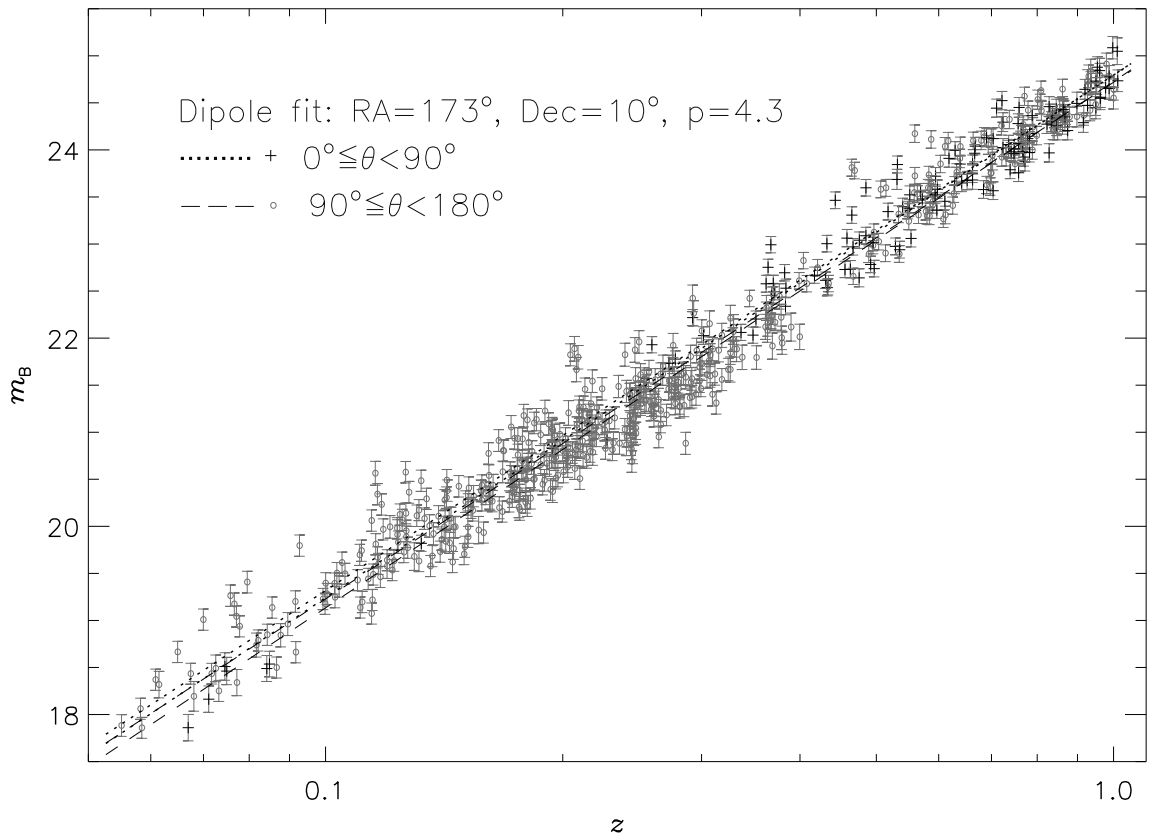}
\caption{The value derived for observer's peculiar velocity from the $m_{\rm B}-z$ plot of SNe~Ia is along RA=173$^\circ$, Dec=10$^\circ$ with $p=4.3$. The upper dotted line represents the fit to the heliocentric SNe~Ia data in $\Sigma_1$ hemisphere, while the lower dashed line represents the same for the $\Sigma_2$ hemisphere. However, both data, after making the corrections for the derived peculiar velocity $v=4.3\times 370$ km s$^{-1}$, are shown as plus (+) and circle (o) symbols in the plot, and when separately fitted, the dotted and dashed lines now almost coincide in the middle.
}
\end{figure*}

\begin{figure*}
\includegraphics[width=15cm]{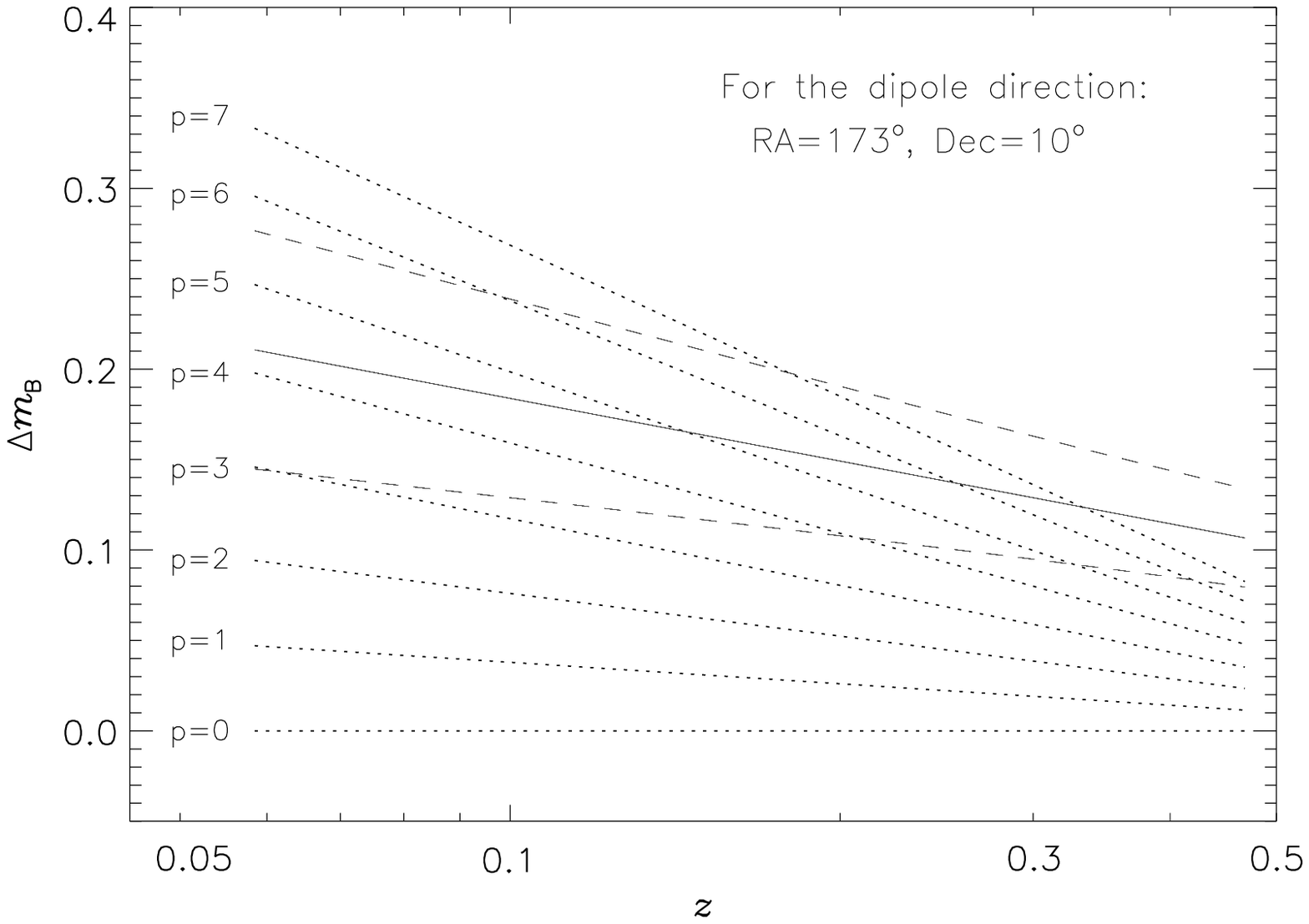}
\caption{For observer's peculiar velocity direction along RA=173$^\circ$, Dec=10$^\circ$, the expected differences $\Delta m_{\rm z}$ between average magnitudes of sources from the two hemispheres, $\Sigma_1$ and $\Sigma_2$, plotted as dotted lines, for mock peculiar velocity values ($p=0$ to 7). 
The unbroken line shows a fit to the actual observed difference at different redshifts in the average magnitudes of sources from the two hemispheres, while the dashed lines above and below the unbroken line represent the $1 \sigma$ uncertainties in the fit. 
}
\end{figure*}


In order to be more certain that there are no effects of some skew distribution in our sample, we divided our JLA sample into two equal halves by picking the odd numbered or even numbered sources in our list and then applying the above procedures separately for each of the two sub-samples. We got quite consistent results in each of these two sub-samples. The values we obtained for the peculiar velocity for the two sub-samples respectively were $p=4.4\pm 1.7$ along RA$=182^{\circ}\pm 17^{\circ}$, Dec$=15^{\circ}\pm11^{\circ}$, and $p=4.3\pm 1.9$ along RA$=157^{\circ}\pm 21^{\circ}$, Dec$=11^{\circ}\pm15^{\circ}$. The difference in the pole positions and $p$ values are well within the quoted statistical uncertainties.

In order to compare directions of the dipoles determined from different datasets, we show  in Fig.~11 the relative positions of the 
estimated directions of various dipoles in the sky. The position of~the pole determined from the Hubble diagram 
of our SNe Ia sample, indicated by $S$, is shown along with the error ellipse. Also shown are the pole positions for other AGN dipoles, 
along with their error ellipses: $N$ (NVSS) (Singal 2011), $T$ (TGSS) (Singal 2019a), $Z$ (DR12Q) (Singal 2018b),  $M$ (MIRAGN) (Singal 2021a,b). The~CMBR pole, at RA$=168^{\circ}$, Dec$=-7^{\circ}$, indicated by $\odot$, has 
negligible errors (Hinshaw, et al. 2009; Aghanim et al. 2020).
It seems that the directions of the dipoles from the Hubble diagram of the SNe Ia sample as well as of mid-infrared 
quasars lie within $2\sigma$ of the CMBR pole, but each of the other three dipoles lies within $\sim 1\sigma$ of the CMBR pole. From that 
we can surmise that the various dipoles, including the CMBR~dipole, are all pointing along the same direction.
Nevertheless, as we pointed out earlier, all these other dipoles have much larger amplitudes than the CMBR dipole, with almost an order 
of magnitude spread, even though various dipole directions in the sky may be lying parallel to each other.
From various dipoles we cannot arrive at a single coherent picture of the solar peculiar velocity, which,   
defined as a motion relative to the local comoving coordinates  and from the CP, a motion with respect to an average universe, 
should after all not depend upon the exact method used for its determination. 

The technique employed here for estimating the peculiar motion of the observer from the Hubble diagram has certain distinct advantages as compared to the elsewhere applied, alternate methods.
The previous methods to determine the peculiar motion of the Solar system from number counts or sky brightness (Singal 2011;  Gibelyou \& Huterer 2012; Rubart \& Schwarz 2013; Tiwari et al. 2015; Colin et al. 2017; Bengaly et al. 2018; Singal 2019a,b; Siewert et al. 2021; Secrest et al. 2021; Singal 2021a,b), 
where aberration played a prominent role, required samples comprising large numbers ($\stackrel{>}{_{\sim}} 10^5$) of distant sources. Moreover, in these techniques a uniform coverage of the whole sky (or at least a large fraction of it), with at most only a small number of gaps, was quite essential, as the techniques depended upon comparing the relative number densities of sources in different parts of the sky.

However the present technique of employing the magnitude-redshift Hubble diagram to estimate the peculiar motion of the observer, does not compare the relative numbers of sources in different regions of the sky. Therefore it does not depend upon a completeness of the survey, nor does it get much affected if the data is combined from a heterogeneous set of various sub-samples. Even a uniform coverage of the whole sky is not essential, all that one requires is that no systematic errors have observationally entered in the redshift and magnitude estimates of individual sources depending upon the direction in sky.  
We have been able to successfully apply this technique to determine the peculiar motion of the solar system from the magnitude-redshift Hubble diagram, comprising a much smaller number of ($\stackrel{<}{_{\sim}} 10^3$) SNe~Ia, because of the presence of a very tight $m_{\rm B}-z$ relation. 

A factor of  $\stackrel{>}{_{\sim}}4$ in the peculiar velocity derived here from SNe IA as compared to the CMBR value, as well as the earlier derived large factors of two to ten in the dipole amplitudes from various AGN datasets 
(Singal 2011; Rubart \& Schwarz 2013; Tiwari et al. 2014; Colin et al. 2017; Bengaly et al. 2018; Singal 2019a,b; Siewert et al. 2021; Secrest et al. 2020; Singal 2021a,b)  
may perhaps be pointers to the need for some rethinking on the conventional interpretation of these dipoles, especially, whether these dipoles do pertain to the peculiar motion of the Solar system. At the same time, an alignment of dipole directions in all cases cannot be fortuitous and is perhaps an indication of a preferred direction (an axis!) in the cosmos, which would imply a breakdown of the cosmological principle, the basic foundation over which the edifice of the modern cosmology has been erected. 

Among the suggestions, inconsistent with the standard model, is a dark flow model implying the existence of a primordial CMBR dipole of non-kinematic origin (Atrio-Barandela et al. 2015).
There seems to be a significant variation in the Hubble parameter $H_0$, depending upon whether it is inferred from an early Universe data based on the flat $\Lambda$CDM model (Aghanim et al. 2020) or the determination is  made from the late Universe based on the distance ladder (Riess et al. 2019,21). A significant Hubble tension is inconsistent with the standard  $\Lambda$CDM model (see also Efstathiou 2021). Although no evidence for anisotropic cosmological expansion from Type Ia supernovae (SNe Ia) is found by Rahman et al. (2021), there may be some evidence that the Hubble parameter is higher in the direction of the CMBR dipole at higher redshifts ($z\sim 1$), which may be a symptom of a deeper cosmological malaise (Krishnan et al. 2021,22; Luongo et al 2022). On the other hand, Horstmann, Pietschke \& Schwarz (2021) determined the solar peculiar motion from the Pantheon sample of SNe Ia to be even smaller than that from the CMBR, though again in the same direction. 
Recently a new approach to disentangle the intrinsic dipole and the kinematic dipole due to our peculiar velocity has been put forward (Nadolny et al. 2021).
\begin{figure*}
\includegraphics[width=15cm]{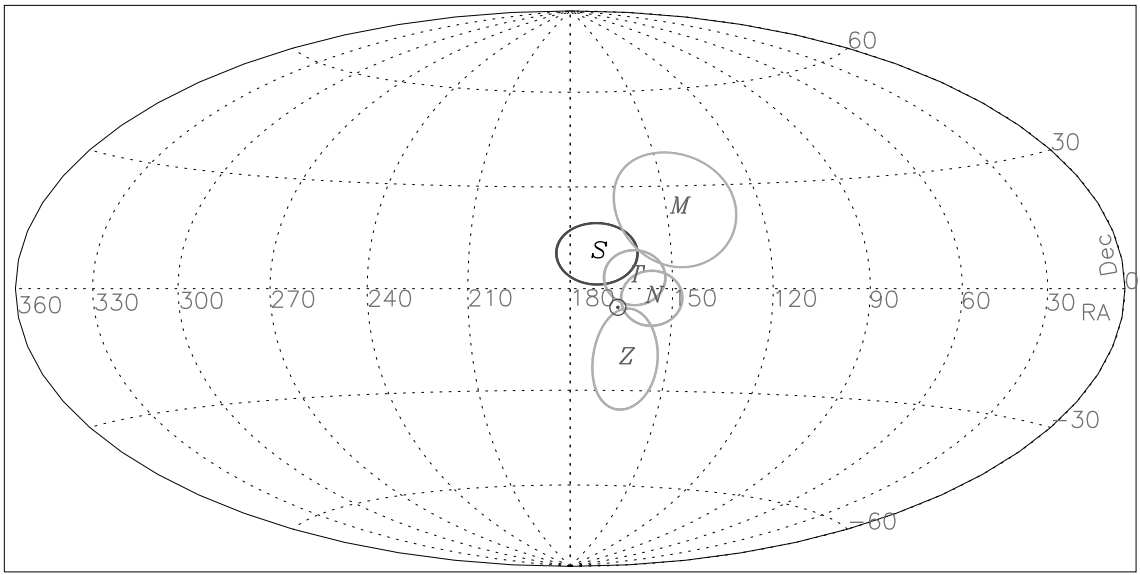}
\caption{The sky, in~the Hammer--Aitoff equal-area projection, showing in equatorial coordinates RA and Dec the position $S$ of~the pole determined from the Hubble diagram of our SNe Ia sample,  along with the error ellipse. Also shown on the map are the other pole positions for various dipoles along with their error ellipses,  $N$ (NVSS), $T$ (TGSS), $Z$ (DR12Q),  $M$ (MIRAGN). The~CMBR pole, at RA$=168^{\circ}$, Dec$=-7^{\circ}$, indicated by $\odot$, has negligible errors.}
\end{figure*}

The cosmological parameters from SNe~Ia data have been determined in the literature 
(Riess et al. 1999; Perlmutter et al. 1999; Betoule et al. 2014; Nielsen, Guffanti \& S. Sarkar 2016; Rubin \& Hayden 2016; Jones et al. 2018; Scolnic et al. 2018; Colin et. al. 2019; Rubin \&  Heitlauf 2020), 
using a peculiar velocity of the observer, as derived from the CMBR dipole, to reduce the heliocentric redshifts to the comoving reference frame. However, a larger peculiar motion would necessitate a fresh look at these determinations, more so as the fits to the $m_{\rm B}-z$ plots vis-\`a-vis theoretical curves of various cosmological models get anchored at lower redshifts, where the displacement  due to the peculiar motion may be substantial. This will be particularly true if the sources in the sample being considered are disproportionally larger in one of the two hemisphere, as it will cause a differential shift at different redshifts in the accordingly observed $m_{\rm B}-z$ plot. Then adjusting a theoretical curve for a cosmological model to match with the observed $m_{\rm B}-z$ plot at low redshifts ($z\stackrel{<}{_{\sim}}  0.1$), might result, in turn, in a relative shift at high redshifts ($z \sim 1$).

More important, even the interpretation of the Hubble plot of SNe~Ia and its comparisons with various cosmological models, suggesting an accelerating Hubble expansion 
(Riess et al. 1999; Perlmutter et al. 1999; Betoule et al. 2014;  Rubin \& Hayden 2016; Jones et al. 2018; Scolnic et al. 2018; Rubin \&  Heitlauf 2020), 
is based on the underlying assumption of the cosmological principle. Thus any doubts on the  cosmological principle will lead to similar doubts in the conventional interpretation of the magnitude-redshift diagram of SNe~Ia to estimate cosmological parameters and the inferences drawn in these as well as in most other important cosmological conclusions.
\section*{Data Availability}
The data underlying this article are available in VizieR Astronomical Server in the public domain at http://vizier.u-strasbg.fr/viz-bin/VizieR. The dataset is downloadable by selecting catalog: J/A+A/568/A22/tablef3. 
\section*{Declarations}
The author has no conflicts of interest/competing interests to declare that are relevant to the content of this article. No funds, grants, or other support of any kind was received from anywhere for this research.

\end{document}